\documentstyle[aps]{revtex}
\begin{document}
\draft
\title{Integrable cases of gravitating static isothermal fluid spheres}
\author{B.V.Ivanov}
\address{Institute for Nuclear Research and Nuclear Energy,\\
Tzarigradsko Shausse 72, Sofia 1784, Bulgaria}
\maketitle

\begin{abstract}
It is shown that different approaches towards the solution of the Einstein
equations for a static spherically symmetric perfect fluid with a $\gamma $%
-law equation of state lead to an Abel differential equation of the second
kind. Its only integrable cases at present are flat spacetime, de Sitter
solution and its Buchdahl transform, Einstein static universe and the
Klein-Tolman solution.
\end{abstract}


\section{Introduction}

The Einstein equations for static and spherically symmetric perfect fluids
have been investigated by many authors \cite{one},\cite{two},\cite{three}.
An abundance of solutions has been found when no equation of state is
prescribed, since then the unknown functions are more than the equations.
The problem becomes very difficult when the fluid's pressure $p\left(
r\right) $ is defined as a known function of the fluid's density $\rho
\left( r\right) $, both depending only on the radial coordinate $r$. A
realistic equation of state is provided by the Newtonian polytropes $p=\frac 
1n\rho ^{1+1/k}$ which have been studied for many years \cite{four},\cite
{five}. In the limit when $k\rightarrow \infty $ a softer isothermal
equation of state emerges 
\begin{equation}
\rho =np,  \label{1}
\end{equation}
usually called the $\gamma $-law, because it is traditionally written as $%
p=\left( \gamma -1\right) \rho $. Physically realistic perfect fluid
solutions have $1\leq n\leq \infty $. Important special cases include dust ($%
n=\infty $), incoherent radiation ($n=3$) and stiff fluid ($n=1$) where the
speed of sound equals the speed of light.

Even for such a simple linear relation, few analytic solutions have been
found. When $n=\infty $ and $\rho \neq 0$ the pressure vanishes, giving the
case of dust. In fact, the density also vanishes and what remains is trivial
flat spacetime. When $n=-1$ the pressure and the density are constant and
the solution is equivalent to a vacuum solution with a cosmological constant
found by de Sitter. We show in this paper that the case $n=-5$ is connected
to it by a Buchdahl transformation \cite{six},\cite{seven},\cite{eight} and
is also soluble. The case $n=-3$ leads to the Einstein static universe \cite
{two}. It is clear that the unphysical cases ($n<1$) should be studied too
since they produce cosmological solutions or are connected to the physical
ones by a general transformation, valid also when no equation of state is
prescribed.

A simple solution for a general $n$ was hidden among the solutions found by
Tolman \cite{nine}. In fact, he studied the field equations under
simplifying assumptions for the metric components, without imposing an
equation of state. The pressure and the density in some of his solutions
happen to satisfy Eq. (1) when certain constants are sent to zero or to
infinity \cite{three}. The first who systematically investigated relation
(1) was O. Klein. He rediscovered the Tolman solution first for $n=3$ and
later for arbitrary $n$ \cite{ten},\cite{eleven}. His approach and results,
published in a not-readily available journal, remained unnoticed for a long
time. Even in Ref. \cite{one} his second work is mentioned as referring to
the polytropic equation of state, which is true only for its beginning. For
a third time this Klein-Tolman (KT) solution was found by Misner and
Zapolsky \cite{twelve}. The radial dependence of the density, however, was
omitted due to a misprint, causing additional confusion. Nevertheless, this
paper became a standard reference towards which further rediscoveries \cite
{thirteen},\cite{fourteen} were directed \cite{fifteen}. Klein also found
numerically a regular solution, starting in phase space from Minkowski
spacetime and spiralling towards the KT solution. His work was based on a
certain first order differential equation. This solution was also
rediscovered later by numeric studies of two and three-dimensional
autonomous dynamical systems \cite{five},\cite{sixteen},\cite{seventeen},%
\cite{eighteen} and further analysed.

The question whether the regular solution, parameterized by $n$, has an
explicit expression has never been answered in a satisfactory way. The
persistent closure of analytic methods on the irregular KT solution suggests
that there are no more integrable cases except it and the few solutions
mentioned above. On the other hand, similar problems have been completely
solved explicitly. For example, static dust solutions are not possible. The
dust must be non-static or charged. In both cases the general solution has
been found \cite{nineteen},\cite{twenty},\cite{twentyone}. The cylindrically
symmetric static case has been solved in simple functions, using two
different gauges \cite{twentytwo},\cite{twentythree}. The relation between
them was clarified in Ref. \cite{twentyfour}. The planar case follows easily
from the cylindrical one \cite{twentythree} or from the spherical one \cite
{seventeen},\cite{twentyfive}.

The purpose of this paper is to derive the integrable cases in a unified
manner and to elucidate the mathematical difficulty of the problem. We show
that in its heart stands the Abel equation of the second kind, whose normal
form is 
\begin{equation}
ww_z-w=f\left( z\right) .  \label{2}
\end{equation}
Its integrable cases depend on the shape of $f\left( z\right) $ and are
tabulated in Ref. \cite{twentysix}. The functions that emerge are in general
transcendental, but simplify for special values of $n$. The integrable cases
that we find are given by $n=-5,-3,-1,\infty $ and the KT solution with $n$
outside the interval $\left( -5.83,-0.17\right) $.

In the following three sections three different approaches are discussed
which invariably lead to Eq. (2) with functions $f\left( z\right) $
possessing the same general structure but with different coefficients. In
Sec. II we start from the well-known Tolman-Oppenheimer-Volkoff (TOV)
equation \cite{nine},\cite{twentyseven} written in a general spherical
metric. In Sec. III the starting point is a differential equation derived by
Klein in curvature coordinates. In Sec. IV we utilize the approach of Haggag
and Hajj-Boutros (HH) described in Ref. \cite{fourteen} for a stiff perfect
fluid in isotropic coordinates. Sec. V is dedicated to the Buchdahl
transformation which supplies the integrable case $n=-5$. Finally, Sec. VI
contains some discussion and conclusions.

\section{Analysis of the TOV equation}

The metric of a static spherically symmetric spacetime reads \cite{three},%
\cite{twentyeight} 
\begin{equation}
ds^2=e^{2\nu }dt^2-e^\lambda dr^2-R^2d\Omega ^2,  \label{3}
\end{equation}
where $d\Omega ^2$ is the metric on the two-sphere and $\nu ,\lambda ,R$
depend on $r$. The Einstein equations are written as 
\begin{equation}
\rho =\frac 1{R^2}-e^{-\lambda }\left( \frac{2R^{\prime \prime }}R+\frac{%
R^{\prime 2}}{R^2}\right) -\left( e^{-\lambda }\right) ^{\prime }\frac{%
R^{\prime }}R,  \label{4}
\end{equation}
\begin{equation}
p=-\frac 1{R^2}+e^{-\lambda }\frac{R^{\prime }}R\left( \frac{R^{\prime }}R%
+2\nu ^{\prime }\right) ,  \label{5}
\end{equation}
\begin{equation}
p=e^{-\lambda }\left[ \frac{R^{\prime \prime }}R+\nu ^{\prime \prime
}+\left( \nu ^{\prime }-\frac{\lambda ^{\prime }}2\right) \left( \nu
^{\prime }+\frac{R^{\prime }}R\right) \right] ,  \label{6}
\end{equation}
where the prime means a derivative with respect to $r$ and units are used
with $8\pi G=c=1$. The contracted Bianchi identity 
\begin{equation}
p^{\prime }=-\left( \rho +p\right) \nu ^{\prime },  \label{7}
\end{equation}
follows from Eqs. (4)-(6) and usually replaces Eq. (6). Thus Eqs. (4), (5)
and (7) determine $\nu ,\lambda ,R,\rho $ and $p$. One of the metric
functions is redundant and can be used to fix the gauge. Different
coordinate systems have been introduced in the literature. Curvature
coordinates (called also Schwarzschild coordinates) are obtained when $R=r$.
Isotropic coordinates have $R=re^{\lambda /2}$ while polar gaussian
coordinates set $e^\lambda =1$. Other coordinates are known as well. Fixing
the gauge and the equation of state equals the number of unknowns and
equations. For the time being, we proceed in full generality to derive the
TOV equation. Let us define the so-called mass function $m\left( r\right) $: 
\begin{equation}
m\left( r\right) \equiv \frac 12R\left( 1-e^{-\lambda }R^{\prime 2}\right) .
\label{8}
\end{equation}
Then Eq. (4) may be written as 
\begin{equation}
\rho =\frac{2m^{\prime }}{R^2R^{\prime }},  \label{9}
\end{equation}
which integrates to 
\begin{equation}
m\left( R\right) =\frac 12\int_0^R\rho R^2dR.  \label{10}
\end{equation}
Passing from $r$ to $R$ dependence, inserting $e^{-\lambda }$ from Eq. (8)
and $\nu _R$ from Eq. (7) into Eq. (5) results in the general TOV equation 
\begin{equation}
p_R=-\frac{\left( \rho +p\right) \left( 2m+pR^3\right) }{2R\left(
R-2m\right) }.  \label{11}
\end{equation}

In the case of dust $p=0$. Then Eq. (11) yields $\rho m=0$, which combined
with Eqs. (9)-(10) gives $\rho =0$. There is no matter and the solution is
trivial flat spacetime. In the general case when $p=p\left( \rho \right) $,
Eq. (9) shows that Eq. (11) is a differential equation for $m\left( R\right) 
$. Let us now introduce the variables $M=m/R,$ $D=\frac 12\rho R^2$ and $P=%
\frac 12pR^2$. Then Eq. (11) becomes 
\begin{equation}
RD_R=2D-\frac{\left( D+P\right) \left( M+P\right) }{\left( 1-2M\right)
p_\rho }.  \label{12}
\end{equation}
Specializing to the $\gamma $-law equation of state and introducing $\tau
=\ln R$ we obtain from Eqs. (12) and (9) the autonomous system 
\begin{equation}
\left( 2M-1\right) D_\tau =D\left[ \left( n+5\right) M-2+\frac{n+1}nD\right]
,  \label{13}
\end{equation}
\begin{equation}
D=M_\tau +M.  \label{14}
\end{equation}
The case $n=0$ is excluded because it gives vanishing density. This system
was derived in polar gaussian coordinates by Collins \cite{five} and further
analysed by him in Refs. \cite{seventeen}, \cite{twentynine}.

Now, let us insert Eq. (14) into Eq. (13) and define $x=M-1/2$. The result
is 
\begin{equation}
xx_{\tau \tau }=\frac{n+1}{2n}x_\tau ^2+\left[ \frac{\left( n+1\right)
\left( n+2\right) +2n}{2n}x+\frac{\left( n+1\right) \left( n+2\right) }{4n}%
\right] x_\tau +A\left( x\right) ,  \label{15}
\end{equation}
\begin{equation}
A\left( x\right) \equiv \left( x+\frac 12\right) \left[ \frac{\left(
n+1\right) ^2+4n}{2n}\left( x+\frac 12\right) -1\right] .  \label{16}
\end{equation}
Eq. (8) indicates that $x<0$. The solution of Eq. (15) determines $M$ and
consequently $m$ as functions of $R$. Then $D$ is given by Eq. (14) which
determines respectively $\rho \left( R\right) $. The pressure is given by
Eq. (1), while $\nu $ follows from Eq. (7) written as 
\begin{equation}
\left( \rho +p\right) \nu _R=-p_R.  \label{17}
\end{equation}
Finally, $e^{-\lambda }R^{\prime 2}$ is found from Eq. (5), written as 
\begin{equation}
e^{-\lambda }R^{\prime 2}=\frac{1+pR^2}{1+2R\nu _R}.  \label{18}
\end{equation}
In order to determine $\lambda \left( r\right) $ one should specify $R\left(
r\right) $, the most simple choice being $R=r$. In polar gaussian
coordinates Eq. (18) yields an equation for $R\left( r\right) $ with
separated variables. The same is true in isotropic coordinates.

Eq. (15) simplifies enormously when $x$ is constant, becoming 
\begin{equation}
M\left[ \frac{\left( n+1\right) ^2+4n}{2n}M-1\right] =0.  \label{19}
\end{equation}
The choice $M_1=0$ leads to $m=p=\rho =0$, i.e. to flat spacetime. The
choice 
\begin{equation}
M_2=\frac{2n}{\left( n+1\right) ^2+4n},  \label{20}
\end{equation}
gives via Eq. (14) $D=M=const$ and $\lambda =const$, $\rho =2D/R^2$ which is
exactly the KT solution. At the centre the density and the pressure have
poles and diverge. This solution does not exist when $-5.83=-3-2\sqrt{2}%
<n<-3+2\sqrt{2}=-0.17$ because then $x_2=M_2-1/2$ is positive or vanishes
(when $n=-1$). In the intervals $-0.17<n<0$ and $n<-5.83$ the solution
exists but $\rho $ and $p$ have different signs.

When $x_\tau \neq 0$ we can decrease the order of Eq. (15) by the standard
change of variables $x_\tau \equiv -y\left( x\right) $: 
\begin{equation}
xyy_x=\frac{n+1}{2n}y^2-\left[ \frac{\left( n+1\right) \left( n+2\right) +2n%
}{2n}x+\frac{\left( n+1\right) \left( n+2\right) }{4n}\right] y+A\left(
x\right) .  \label{21}
\end{equation}
Eq. (21) falls in the class 
\begin{equation}
\left[ g_1\left( x\right) y+g_0\left( x\right) \right] y_x=f_2\left(
x\right) y^2+f_1\left( x\right) y+f_0\left( x\right) .  \label{22}
\end{equation}
There is a standard procedure for the solution of such equations \cite
{twentysix}. It consists of two changes of variables which bring them to the
Abel equations of the second kind given by Eq. (2) or by 
\begin{equation}
ww_\zeta =g\left( \zeta \right) w+1.  \label{23}
\end{equation}
This procedure is much easier when $g_0=0$ and $f_2=const$ as in Eq. (21).
Namely, we have $w=yE$, $f\left( z\right) =f_0E/f_1$, $g\left( \zeta \right)
=f_1/f_0E$ and 
\begin{equation}
E=\exp \left( -\int \frac{f_2}{g_1}dx\right) ,  \label{24}
\end{equation}
\begin{equation}
z\left( x\right) =\int \frac{f_1}{g_1}Edx,  \label{25}
\end{equation}
\begin{equation}
\zeta \left( x\right) =\int \frac{f_0}{g_1}E^2dx.  \label{26}
\end{equation}
Since $f_1$ is a simpler polynomial than $f_0$ we shall use Eqs. (2) and
(25). The use of Eqs. (23) and (26) is more complicated, but does not bring
additional integrable cases.

Applied to Eq. (21) the chosen alternative of the general method yields 
\begin{equation}
w=yx^{-\frac{n+1}n},  \label{27}
\end{equation}
\begin{equation}
z=-\frac 1{4n}\int \left[ \left( n^2+5n+2\right) 2x+\left( n+1\right) \left(
n+2\right) \right] x^{-\frac{n+1}{2n}-1}dx.  \label{28}
\end{equation}
The case $n=1$ (stiff fluid) is special because a logarithmic term appears
in $z$%
\begin{equation}
z=-4\ln |x|+\frac 3{2x},  \label{29}
\end{equation}
\begin{equation}
f\left( z\right) =-\frac{\left( 2x+1\right) \left( 4x+1\right) }{x\left(
8x+3\right) }.  \label{30}
\end{equation}
The relation $z\left( x\right) $ is transcendental and throws $f\left(
z\right) $ out of the tables with integrable cases present in Ref. \cite
{twentysix}. The case $n=-1$ also leads to a logarithmic term but its
coefficient vanishes. This case is integrable. In fact, we may go back
directly to Eq. (13) which becomes $D_\tau =2D$ if $M\neq 1/2$ and yields $%
\rho =-p=const$. This is the well-known de Sitter solution. When $M=1/2$ Eq.
(13) is satisfied identically and we obtain formally the KT solution (20)
with $n=-1$, but it has $e^{-\lambda }R^{\prime 2}=0$ which is unacceptable.

In the generic case $n\neq \pm 1$ and Eq. (28) integrates to 
\begin{equation}
z=-\frac{x^{-\frac{n+1}{2n}}}{2\left( n-1\right) }\left[ \left(
n^2+5n+2\right) 2x-\left( n-1\right) \left( n+2\right) \right] ,  \label{31}
\end{equation}
\begin{equation}
f\left( z\right) =\frac{\left( n-1\right) \left( 2x+1\right) \left[ \left(
n^2+6n+1\right) 2x+\left( n+1\right) ^2\right] z}{\left[ \left(
n^2+5n+2\right) 2x-\left( n-1\right) \left( n+2\right) \right] \left[ \left(
n^2+5n+2\right) 2x+\left( n+1\right) \left( n+2\right) \right] }.  \label{32}
\end{equation}
The $z$-$x$ connection in Eq. (31) is transcendental, except for special
values of $n$, and Eq. (2) is non-integrable in general. Let us investigate
the special cases.

When $n=-3$, $f_1$ divides $f_0$ and $f\left( z\right) =8z$, $z=-\frac 12%
\left( 2x+1\right) x^{-1/3}$. This is an integrable case, corresponding to
the Einstein static universe. It is discussed in more details in the next
section. There is also the formal $n=-3$ case of the KT solution (20). It
does not exist since $x=1/4>0$.

One may try to simplify Eq. (31) by nullifying the coefficients on the
right. The condition $n^2+5n+2=0$ gives $n=\frac 12\left( -5\pm \sqrt{17}%
\right) $. This is of no good since the radical enters the power of $x$. The
other possibility $n=-2$ looks more promising. Then Eq. (31) becomes $%
x^{3/4}=-3z/4$ and Eq. (32) reduces to 
\begin{equation}
f\left( z\right) =\frac{21}{16}z-\frac 9{16}\left( \frac 43\right)
^{4/3}z^{-1/3}+\frac 3{64}\left( \frac 43\right) ^{8/3}z^{-5/3}.  \label{33}
\end{equation}
This function leads to an integrable equation when the coefficient in front
of $z$ is $-3/16$ which is not true here.

There are several $n$ which convert Eq. (31) into an algebraic equation for $%
x$ of fourth order or lower. It can be solved explicitly for $x\left(
z\right) $ and the answer replaced in Eq. (32). Third and fourth order
equations appear when $n=\pm 1/5,\pm 1/7,-1/2,-3/5$. The radical structure
of $f\left( z\right) $, however, is incompatible with the tables with
integrable $f\left( z\right) $. Second order equations appear when $n=\mp
1/3 $ respectively: 
\begin{equation}
4x=-5\pm \sqrt{25+48z},  \label{34}
\end{equation}
\begin{equation}
6zx=17\pm \sqrt{17+42z}.  \label{35}
\end{equation}
Unfortunately, the only integrable functions with square roots include the
term $\sqrt{z^2+z_0}$ which is not present in the above relations.

Comparison between Eqs. (31)-(32) and Eq. (20) shows how complex must be the
numerical regular solution, which starts in $x,y$ coordinates from flat
spacetime and focuses on the KT solution, following a spiral around it. The
innocent parameter $n$, introduced in Eq. (1), proliferates like cancer in
the process of solution, ending with the intricate coefficients in $z\left(
x\right) $ and $f\left( z\right) $. It even determines the transcendental or
algebraic nature of $z\left( x\right) $. In conclusion, the only integrable
cases found within this approach are $n=\infty $ (trivial dust solution), $%
n=-1$ (de Sitter solution), $n=-3$ (Einstein static universe) and $M=const$
(the KT solution).

Finally, let us discuss for comparison the case of planar symmetry, which is
solvable. Going to polar gaussian coordinates, the metric element reads 
\begin{equation}
ds^2=e^{2\nu }dt^2-dr^2-R^2\left( dx_2^2+F\left( x_2\right) ^2dx_3^2\right) ,
\label{36}
\end{equation}
where $F\left( x_2\right) =\sin x_2$ for spherical symmetry and $F\left(
x_2\right) =1$ for planar symmetry \cite{seventeen},\cite{twentyfive}. It is
possible to generalize the TOV equation to encompass both cases. Instead of
Eq. (13) one should write 
\begin{equation}
\left( 2M-K\right) D_\tau =D\left[ -2K+\left( n+5\right) M+\frac{n+1}n%
D\right] ,  \label{37}
\end{equation}
where $K=1$ or $0$, corresponding to spherical or planar symmetry
respectively. In the second case Eq. (37) simplifies 
\begin{equation}
2MM_{\tau \tau }=\frac{n+1}nM_\tau ^2+bMM_\tau +aM^2,  \label{38}
\end{equation}
where $a=n+1/n+6$, $b=n+2/n+5$. This is the analog of Eq. (15). Proceeding
in the same way we obtain again the Abel equation (2) with $y=-M_\tau $ and 
\begin{equation}
w=yM^{-\frac{n+1}{2n}},  \label{39}
\end{equation}
\begin{equation}
z=\frac{bn}{1-n}M^{\frac{n-1}{2n}},  \label{40}
\end{equation}
\begin{equation}
f\left( z\right) =\frac{\left( n-1\right) a}{nb^2}z.  \label{41}
\end{equation}
As mentioned before, Eq. (2) with $f\left( z\right) =\alpha z+\beta ,$ where 
$\alpha $ and $\beta $ are constants, is integrable. The solution, in
parametric form, reads 
\begin{equation}
z=Ce^T-\frac \beta \alpha ,  \label{42}
\end{equation}
\begin{equation}
w=C\sigma e^T,  \label{43}
\end{equation}
\begin{equation}
T=-\int \frac{\sigma d\sigma }{\sigma ^2-\sigma -\alpha },  \label{44}
\end{equation}
with $C$ being an arbitrary constant. In Ref. \cite{seventeen} the problem
was solved in a different way, by introducing the variable $\tilde D=D/M$.
Then Eqs. (14) and (37) are equivalent to 
\begin{equation}
2\tilde D_\tau =\tilde D\left( n+7+\frac{1-n}n\tilde D\right) .  \label{45}
\end{equation}
This is a Bernoulli equation and is easily solved. Further details may be
found in Ref. \cite{seventeen} where also a connection with earlier work 
\cite{thirty},\cite{thirtyone} on the particular cases $n=1$ and $n=3$ is
established.

\section{The approach of Klein}

This approach was developed in curvature coordinates where Eqs. (4), (5) and
(7) simplify to 
\begin{equation}
p=\frac 2r\nu ^{\prime }e^{-\lambda }-\frac 1{r^2}\left( 1-e^{-\lambda
}\right) ,  \label{46}
\end{equation}
\begin{equation}
\rho =\frac 1r\lambda ^{\prime }e^{-\lambda }+\frac 1{r^2}\left(
1-e^{-\lambda }\right) ,  \label{47}
\end{equation}
\begin{equation}
p^{\prime }=-\left( \rho +p\right) \nu ^{\prime }.  \label{48}
\end{equation}
Imposing the $\gamma $-law, we can integrate Eq. (48): 
\begin{equation}
p=p_0e^{-\left( n+1\right) \nu }.  \label{49}
\end{equation}
Like before, the case of dust $n=\infty $, $p=0$ gives a trivial solution
with constant $\nu $ and $\lambda $. Supposing that $p\neq 0$, let us
introduce the variables $s=p_0r^2$, $\xi =e^{\left( n+1\right) \nu }/s$ and 
\begin{equation}
x=\xi e^{-\lambda }=\frac{e^{-\lambda }}{pr^2}.  \label{50}
\end{equation}
Obviously $x$ and $\xi $ always have the same sign and are positive when the
pressure is positive.

We shall derive an equation for $x$, similar to Eq. (15). In Sec. II, III
and IV the main functions will be denoted by $x$ although they are
different, in order to simplify notation and to stress the role of Eq. (22)
in the whole problem. Let us multiply Eq. (46) by $\left( n+1\right) /4$,
Eq. (47) by $-1/2$ and sum. The result is 
\begin{equation}
4\left( sx\right) _s-\left( n+3\right) \left( 1-e^{-\lambda }\right) \xi
=1-n.  \label{51}
\end{equation}
Let us introduce $2\tau =\ln s$. It differs by a constant from the variable
in the previous section, but this is not important since the final equation
will be autonomous. Laying temporarily aside the special case $n=-3$, we can
express $\xi $ from Eq. (51) as 
\begin{equation}
\xi =\frac 1{n+3}\left[ 2x_\tau +\left( n+7\right) x+n-1\right] .  \label{52}
\end{equation}
Eq. (46) may be written in the following way: 
\begin{equation}
4\left( s\xi \right) _s-\left( n+1\right) \left( e^\lambda -1\right) \xi
=\left( n+1\right) e^\lambda .  \label{53}
\end{equation}
The usage of Eqs. (50) and (52) transforms this relation into an autonomous
second-order equation for $x$. The change of variables $x_\tau =-y\left(
x\right) $ applies again, leading to 
\begin{eqnarray}
2\left( n+3\right) xyy_x &=&2\left( n+1\right) y^2+2\left( n+11\right)
xy-\left( n+1\right) \left( 3n+1\right) y-4\left( n+7\right) x^2  \nonumber
\label{54} \\
&&+\left( n^3+9n^2+11n+11\right) x+\left( n+1\right) ^2\left( n-1\right) .
\label{54}
\end{eqnarray}
This is exactly the equation of Klein derived in Refs. \cite{ten},\cite
{eleven}, where the notation $n=2n_{Kl}+1$ and $x=\left( n_{Kl}+1\right)
^2x_{Kl}$ was used. He studied it by series expansion and numerically, and
was the first to find its regular solution (see Fig. (1) from Ref.\cite{ten}%
). When $x\left( \tau \right) $ is known Eq. (52) gives $\xi $, and Eq. (50)
determines both $\lambda $ and $p$. Then Eq. (49) becomes an expression for $%
\nu $ and finally $\rho $ is given by Eq. (1). Modulo coefficients, Eq. (54)
is the same as Eq. (21) and also falls in the class (22). Therefore, the
procedure described in Sec. II can be applied to bring it to the form of Eq.
(2) with 
\begin{equation}
w=yx^{-\frac{n+1}{n+3}},  \label{55}
\end{equation}
\begin{equation}
z=\frac 12\left[ \left( n+11\right) x+3n+1\right] x^{-\frac{n+1}{n+3}},
\label{56}
\end{equation}
\begin{equation}
f\left( z\right) =-\frac{2\left[ 4x-\left( n+1\right) ^2\right] \left[
\left( n+7\right) x+n-1\right] z}{\left[ 2\left( n+11\right) x-\left(
n+1\right) \left( 3n+1\right) \right] \left[ \left( n+11\right)
x+3n+1\right] }.  \label{57}
\end{equation}
There are no logarithmic terms in $z$ within this approach. As with Eq.
(31), $x\left( z\right) $ is transcendental except in special cases.

Eq. (54) shows that the KT solution is given here by 
\begin{equation}
x_0=\frac{\left( n+1\right) ^2}4.  \label{58}
\end{equation}
The other root $x=-\frac{n-1}{n+7}$ leads to $\xi =0$ and $e^{2\nu
}=e^\lambda =0$, $\rho =np=\infty $ which is unacceptable. Now, since $%
x_0\geq 0$ for any $n$, we must ensure that $\xi _0>0$ in order to have
positive $e^{-\lambda }$. Eq. (52) gives $\xi _0=n^2+6n+1$ and we obtain the
same conditions for the existence of the KT solution as in the previous
section.

In the case $n=-1$ Eq. (48) yields $p=p_0$, $\rho =-p_0$, while the sum of
Eqs. (46) and (47) provides the relation $2\nu =-\lambda $. Eq. (47) is a
linear equation for $e^{-\lambda }$, its solution being 
\begin{equation}
e^{-\lambda }=1-\frac{2m_0}r+\frac{p_0}3r^2,  \label{59}
\end{equation}
where $m_0$ is a constant of integration, identified as the gravitating
mass. This is precisely the Kottler solution \cite{two}. It can be used as a
regular interior solution when $m_0=0$. Then it becomes the de Sitter
solution. The KT solution for $n=-1$ has $x_0=0$ and does not exist.

When $n=-3$ Eq. (51) determines directly $x$: $x=1+x_1/s$, while Eq. (49)
gives $p=p_0e^{2\nu }$. Then we have from Eq. (50) $e^{-2\nu -\lambda
}=x_1+s $. Eq. (46) becomes a linear equation for $e^{2\nu }$ when these
results are taken into account. There are two possibilities. When the
constant $x_1=0$, $x=1$, which is the formal KT solution (58) for $n=-3$. In
fact, it does not exist. When $x_1\neq 0$ the solution is 
\begin{equation}
x_1e^{2\nu }=1+C_1\left( 1+\frac{x_1}s\right) ^{1/2},  \label{60}
\end{equation}
where $C_1$ is another integration constant. A regular solution is obtained
when $x_1=1$ and $C_1=0$. Then $\nu =0$ and 
\begin{equation}
e^\lambda =\frac 1{1+p_0r^2}.  \label{61}
\end{equation}
The last two equations represent the metric of the Einstein static universe.
The pressure and the density are constant, $p=p_0$, $\rho =-3p_0$.

There are two cases when the coefficients of $z\left( x\right) $ are
simplified, namely $n=-11$ and $n=-1/3$. In the first case Eq. (57) reads 
\begin{equation}
-320f\left( z\right) =75z+16^{8/5}z^{-3/5}+22\times 16^{4/5}z^{1/5}.
\label{62}
\end{equation}
The function 
\begin{equation}
f\left( z\right) =c_1z+c_2z^{q_1}+c_3z^{q_2},  \label{63}
\end{equation}
is integrable for a set of $\left( q_1,q_2\right) $ but $\left(
-3/5,1/5\right) $ is not among them. When $n=-1/3$ we have 
\begin{equation}
f\left( z\right) =-\frac{15}{64}z+\frac 7{96}\left( \frac{16}3\right)
^{4/3}z^{-1/3}-\frac 1{192}\left( \frac{16}3\right) ^{8/3}z^{-5/3}.
\label{64}
\end{equation}
The set $\left( -1/3,-5/3\right) $ is integrable, but only when $c_1=-3/16$
which is not the case here.

Finally, there are few $n$ when Eq. (56) is an algebraic equation up to the
fourth order. Only the cases $n=-5,-2,1$ are candidates for integrability
because they lead to quadratic equations. Of these, $n=-5,-2$ yield radicals
resembling those in Eqs. (34) and (35) and should be rejected. More
interesting is the case of stiff fluid when 
\begin{equation}
x=\frac 1{72}\left( z^2-24\pm z\sqrt{z^2-48}\right) ,  \label{65}
\end{equation}
\begin{equation}
f\left( z\right) =-\frac{2z\left[ \left( z^2-60\right) x-4\right] }{9\left[
\left( z^2-24\right) x-8\right] }.  \label{66}
\end{equation}
The radicals in $f\left( z\right) $ are of the necessary type, but its
structure is too complex to figure in the tables with integrable cases.

Like in Sec. II, the only integrable cases found are $n=-3,-1,\infty $ and $%
x=x_0$. The unsuccessful candidates for explicit solutions have in general
different values of $n$ in the two approaches. When they coincide, as is the
case $n=1$, the reasons for rejection are different - a logarithmic term in
the TOV-Collins approach and a complicated $f\left( z\right) $ in the Klein
approach.

It is interesting to compare the main variable in this section $x\equiv x_K$
to the variables in the TOV approach, specialized to curvature coordinates.
Eq. (8) becomes 
\begin{equation}
M=\frac mr=\frac 12\left( 1-e^{-\lambda }\right) ,  \label{67}
\end{equation}
and transforms Eq. (50) into 
\begin{equation}
x_K=\frac{1-2M}{2nD}.  \label{68}
\end{equation}
Thus, not only $x=M-1/2$, but also the above combination satisfy Abel
equations of the second kind. The function $x_K$ resembles $\tilde D$, used
in the planar case. It is well-known that when $\tilde D$ satisfies the
Bernoulli Eq. (45), $\tilde D^{-1}=M/D$ satisfies a linear equation. This
fact stresses once more the conclusion that the spherical case is much more
complicated than the planar one.

\section{Approach in isotropic coordinates}

One can pass from arbitrary to isotropic coordinates in Eqs.(4)-(6) by
putting $R=re^{\lambda /2}$. Let us make also the change $s=\ln r$. Then 
\begin{equation}
\rho r^2e^\lambda =-\lambda _{ss}-\frac 14\lambda _s^2-\lambda _s,
\label{69}
\end{equation}
\begin{equation}
pr^2e^\lambda =\frac 14\lambda _s^2+\lambda _s\nu _s+\lambda _s+2\nu _s,
\label{70}
\end{equation}
\begin{equation}
pr^2e^\lambda =\frac 12\lambda _{ss}+\nu _{ss}+\nu _s^2.  \label{71}
\end{equation}
Let us introduce next the variable 
\begin{equation}
\frac 1t=\lambda _s+2,  \label{72}
\end{equation}
and impose the $\gamma $-law equation of state. An expression for $\nu _s$
is obtained from Eqs. (69) and (70): 
\begin{equation}
\nu _s=\frac{t_s}{nt}-\frac{n+1}{4n}\left( \frac 1t-4t\right) .  \label{73}
\end{equation}

Next, let us combine Eqs. (70) and (71) and replace in them $\lambda _s$ and 
$\nu _s$ from the above equations. A long, but straightforward computation
produces an autonomous equation for $t$%
\begin{eqnarray}
2ntt_{ss} &=&-2\left( 1-n\right) t_s^2+\frac 12\left( n^2+5n+2\right)
t_s-2\left( n+1\right) \left( n+2\right) t^2t_s-2\left( n+1\right) ^2t^4 
\nonumber \\
&&\ +\left[ \left( n+1\right) ^2+2n\right] t^2-\frac n2-\frac{\left(
n+1\right) ^2}8  \label{74}
\end{eqnarray}
A solution of this master equation determines all characteristics of the
metric and the fluid.

When $n=1$ Eq. (74) is exactly Eq. (9) from Ref. \cite{fourteen}. In this
section we generalize the HH approach to arbitrary $n$ and bring it to its
logical end - the Abel equation (2). We first lower the order of the
polynomial in Eq. (74) by setting $t^2=x/4$ and then perform the change of
variables 
\begin{equation}
\left( \sqrt{x}\right) _s=-\frac 12y.  \label{75}
\end{equation}
The condition $x\geq 0$ should be maintained throughout the calculations.
Eq. (74) acquires its final form 
\begin{eqnarray}
2nxyy_x &=&\left( n-1\right) y^2+\left[ \left( n+1\right) \left( n+2\right)
x-\left( n^2+5n+2\right) \right] y  \nonumber  \label{76} \\
&&+\left( x-1\right) \left[ 4n+\left( n+1\right) ^2-\left( n+1\right)
^2x\right] .  \label{76}
\end{eqnarray}
It falls in the class (22) and resembles Eqs. (21) and (54), but its
coefficients are different functions of $n$. Proceeding like before, we get 
\begin{equation}
w=yx^{-\frac{n-1}{2n}},  \label{77}
\end{equation}
\begin{equation}
\left( n-1\right) z=\left[ \left( n-1\right) \left( n+2\right)
x+n^2+5n+2\right] x^{-\frac{n-1}{2n}},  \label{78}
\end{equation}
\begin{equation}
f\left( z\right) =\frac{\left( n-1\right) \left( x-1\right) \left[ 4n+\left(
n+1\right) ^2-\left( n+1\right) ^2x\right] z}{\left[ \left( n+1\right)
\left( n+2\right) x-\left( n^2+5n+2\right) \right] \left[ \left( n-1\right)
\left( n+2\right) x+n^2+5n+2\right] },  \label{79}
\end{equation}
in the generic case $n\neq \pm 1$.

The stiff fluid case leads to a logarithmic term in $z\left( x\right) $
which makes 
\begin{equation}
f\left( z\right) =-\frac{2\left( x-1\right) \left( x-2\right) }{3x-4},
\label{80}
\end{equation}
non-integrable.

The case $n=-1$ is pseudo-logarithmic since the coefficient in front of $\ln
x$ vanishes. We have $x=1/z$ and $f\left( z\right) =2z-2$. This case is
integrable and the solution is given by Eqs. (42)-(44) with $\alpha =-\beta
=2$.

The case $n=-3$ leads to a gross simplification of $f\left( z\right) $ and
is also soluble. One obtains $f\left( z\right) =2z$ i.e. $\alpha =2$, $\beta
=0$.

The other candidate cases may be investigated in the same manner as in Secs.
II and III. The values $n=\pm 1/3$ lead to quadratic equations with radicals
of the wrong type. The case $n=-2$ yields 
\begin{equation}
f\left( z\right) =\frac{21}{16}z-\frac{18}{16}\left( \frac 43\right)
^{4/3}z^{-1/3}-\frac 3{16}\left( \frac 43\right) ^{8/3}z^{-5/3}.  \label{81}
\end{equation}
This equation belongs to the class (63) but again $c_1\neq -3/16$ and
integrability is not gained.

Finally, let us discuss the KT solution in the HH approach. When $x=const$
Eq. (76) becomes purely algebraic and has two roots: $x_1=1$ and 
\begin{equation}
x_2=1+\frac{4n}{\left( n+1\right) ^2}.  \label{82}
\end{equation}
When $n=-1$, $x_2$ does not exist. In fact, the requirement $x_2>0$ shows
that $n$ must satisfy the conditions derived in Sec. II. The first root
leads to $t_1=\pm 1/2$. In the first subcase Eqs. (72) and (73) give flat
spacetime. In the second subcase the following line element is obtained 
\begin{equation}
ds^2=dt^2-r^{-4}\left( dr^2+r^2d\Omega ^2\right) .  \label{83}
\end{equation}
The transformation $\tilde r=1/r$ converts this element into the usual
element for flat spacetime.

The second root $x_2$ represents the KT solution in isotropic coordinates,
namely $e^\lambda =r^{\alpha _1}$, $e^{2\nu }=r^{\alpha _2}$ where 
\begin{equation}
\alpha _1=\pm \frac{2\left( n+1\right) }{\sqrt{4n+\left( n+1\right) ^2}}-2,
\label{84}
\end{equation}
\begin{equation}
\alpha _2=\pm \frac 4{\sqrt{4n+\left( n+1\right) ^2}}.  \label{85}
\end{equation}
When $n=1$, $\alpha _1=\pm \sqrt{2}-2$ and $\alpha _2=\pm \sqrt{2}$. These
values were found in Ref. \cite{fourteen}, where the KT solution was
discovered for a fifth time.

The integrable cases found in the HH approach coincide with those found in
the Klein or the TOV-Collins approaches. In order to make connection with
the last one, we must pass in Sec. II to isotropic coordinates. We have 
\begin{equation}
R^{\prime }=e^{\lambda /2}\left( 1+\frac{\lambda _s}2\right) ,  \label{86}
\end{equation}
\begin{equation}
2x=2M-1=-R^{\prime 2}e^{-\lambda }=-\frac 1{4t^2}.  \label{87}
\end{equation}
We cannot obtain Eq. (74) from Eq. (15) by replacing there just $x$ from Eq.
(87) because $s=\ln r$, while $\tau =\ln R=s+\lambda /2$. The necessary
additional relations are 
\begin{equation}
x_\tau =\frac{t_s}{2t^2},  \label{88}
\end{equation}
\begin{equation}
x_{\tau \tau }=\frac{t_{ss}}t-\frac{2t_s^2}{t^2}.  \label{89}
\end{equation}
Therefore, the HH approach combines the TOV equation in isotropic
coordinates together with the linear equation (18) for $\lambda $.

\section{Buchdahl transformation and the case $n=-5$}

This transformation was found by Buchdahl \cite{six} and rediscovered by
Glass and Goldman \cite{seven}. Its general formulation refers to static
perfect fluid solutions, not necessarily satisfying an equation of state. In
the case of spherical symmetry it states that if 
\begin{equation}
ds^2=e^{2\nu }dt^2-e^\lambda \left( dr^2+r^2d\Omega ^2\right) ,  \label{90}
\end{equation}
is the metric of a perfect fluid solution with pressure $p$ and density $%
\rho $ then there is a reciprocal solution which has $\nu _b=-\nu $, $%
\lambda _b=4\nu +\lambda $ and 
\begin{equation}
p_b=e^{-4\nu }p,  \label{91}
\end{equation}
\begin{equation}
\rho _b=-e^{-4\nu }\left( \rho +6p\right) .  \label{92}
\end{equation}
The transformation is simplest in isotropic coordinates and may be applied
to the results obtained in the previous section. When $p$ and $\rho $
satisfy the $\gamma $-law, the same is true for $p_b$ and $\rho _b$ but with
a different parameter: 
\begin{equation}
\rho _b=-\left( n+6\right) p_b.  \label{93}
\end{equation}
Thus, the transformation of the parameter is $n\rightarrow -\left(
n+6\right) $. It is clear that when the starting solution is physically
realistic ($n\geq 1$), the transformed one is unphysical, $-\left(
n+6\right) \leq -7$. However, when the starting solution is unphysical and $%
n\leq -7$, then the reciprocal one is physical. When $n$ is in the interval $%
-7<n<1$ both solutions are unphysical. The existence of this transformation
teaches that unphysical solutions should not be neglected a priori and the
whole spectrum of $n$ must be investigated.

Let us apply the transformation to the integrable cases found in Sec. IV.
The KT solution is self-reciprocal, i.e. transforms into itself. The reason
is that $4n+\left( n+1\right) ^2$ is invariant under the transformation.
Taking $\alpha _{1-}$ and $\alpha _{2-}$ as the powers of $r$ in the
starting solution, it is easy to show that $\alpha _{2+}=-\alpha _{2-\text{ }%
}$ and $\alpha _{1+}=\alpha _{1-}+2\alpha _{2-}$ when the plus solution is
Buchdahl transformed. Solutions with $n$ outside the interval $\left(
-5.83,-0.17\right) $ transform between themselves.

The case $n=-3$ is also self-reciprocal, as was noticed already by Buchdahl.
This explains why $\nu =0$ when the element of the Einstein static universe
is written in isotropic coordinates \cite{thirtytwo}. This is necessary to
ensure the equality between the starting and the transformed solution.

The interesting case is the de Sitter solution ($n=-1$) which transforms
into an explicit solution with $n=-5$. In isotropic coordinates the de
Sitter solution reads \cite{thirtytwo} 
\begin{equation}
e^{2\nu }=\left( \frac{1+cr^2}{1-cr^2}\right) ^2,  \label{94}
\end{equation}
\begin{equation}
e^\lambda =\left( 1-cr^2\right) ^{-2},  \label{95}
\end{equation}
and $p=12/c$, $\rho =-12/c$ where $c$ is some constant. The transformed
solution has 
\begin{equation}
e^{2\nu _b}=\left( \frac{1-cr^2}{1+cr^2}\right) ^2,  \label{96}
\end{equation}
\begin{equation}
e^{\lambda _b}=\frac{\left( 1+cr^2\right) ^4}{\left( 1-cr^2\right) ^6},
\label{97}
\end{equation}
\begin{equation}
p_b=\left( \frac{1-cr^2}{1+cr^2}\right) ^4\frac{12}c,  \label{98}
\end{equation}
and, of course, $\rho _b=-5p_b$. Eq. (72) supplies the corresponding $t$: 
\begin{equation}
t=\frac{1-cr^2}{2\left( 1+cr^2\right) },  \label{99}
\end{equation}
\begin{equation}
t_b=\frac{1-c^2r^4}{2\left( 1+10cr^2+c^2r^4\right) }.  \label{100}
\end{equation}
From Eq. (75) we get $x=4t^2$ and $y=-4rt_r$. When $t_b$ is plugged in these
relations we obtain the parametric solution $x\left( r\right) $, $y\left(
r\right) $ of Eq. (76) for $n=-5$: 
\begin{equation}
xyy_x=\frac 35y^2-\frac 65xy+\frac 15y+\frac 25\left( 4x^2-3x-1\right) .
\label{101}
\end{equation}
Eqs. (78) and (79) yield in this case 
\begin{equation}
-3z=\left( 9x+1\right) x^{-3/5},  \label{102}
\end{equation}
\begin{equation}
f\left( z\right) =\frac{6\left( x-1\right) \left( 4x+1\right) }{\left(
6x-1\right) \left( 9x+1\right) }z.  \label{103}
\end{equation}
Eq. (102) is a fifth order equation for $x$ and seems to be transcendental.
Obviously $f\left( z\right) $ is not among the tabulated integrable
functions. Eq. (101) does not coincide also with any of the equations in
Sec.1.3.4. from Ref. \cite{twentysix} which belong to the class (22). If we
use Eq. (23) instead, $g\left( \zeta \right) =1/f\left( z\right) $ and 
\begin{equation}
3\zeta x^{6/5}=6x^2+18x+1.  \label{104}
\end{equation}
This equation is even more complicated than Eq. (102) and $g\left( \zeta
\right) $ is not among the few integrable cases listed in Sec.1.3.2 from the
same handbook. Unless some mistake has been made, Eq. (101) is integrable,
but is not covered by Ref. \cite{twentysix}. This situation is not unique.
Recently, non-static charged perfect fluid distributions were discovered
which too are missing in the handbooks with solutions of differential
equations \cite{thirtythree}.

\section{Discussion and conclusions}

In this paper we have discussed explicit solutions of the Einstein equations
for a static spherically symmetric perfect fluid with the $\gamma $-law
equation of state. Three approaches may be found in the literature which, at
first sight, have nothing in common. The approach of Collins transforms the
TOV equation in polar gaussian coordinates into a two-dimensional autonomous
system of differential equations. He studies it numerically \cite{five},\cite
{seventeen},\cite{twentynine}. This is a representative example of other
similar dynamical systems \cite{sixteen},\cite{eighteen}. In his approach
Klein finds a second-order autonomous master equation, whose solution
determines all characteristics of the metric and the fluid. He lowers its
order and studies it by series expansion and numerically. Klein also
discovers a simple exact but singular solution. This work is done in
curvature coordinates \cite{ten},\cite{eleven}. Haggag and Hajj-Boutros
perform a similar study but in isotropic coordinates and only for a stiff
fluid \cite{fourteen}. They also obtain an autonomous second-order equation
and transform it into another second-order equation (see Eqs. (9) and (14)
in their paper). Then they look for polynomial solutions and find either
flat spacetime or the KT solution.

We have generalized these three approaches and pursued them further, till
their logical end - an equation, whose integrable cases are tabulated in the
handbooks, like Ref. \cite{twentysix}. Surprisingly, we invariably reach the
Abel equation of the second kind (2). It behaves like a 'strange attractor'
and underlines the common features in the different approaches. The question
of integrability is answered by the form of $f\left( z\right) $, the known
integrable cases being tabulated. For this purpose we have generalized to an
arbitrary coordinate system the dynamical system of Collins and to arbitrary 
$n$ the HH approach . The functions $z\left( x\right) $ and $f\left(
x\right) $ have the same general structure in all approaches but with
different coefficients. The relations between the master variables $x$, $x_K$
and $x_{HH}$ have been elucidated, the TOV-Collins approach serving as a
basis.

The integrable cases found are one and the same and include $n=\infty $
(trivial dust solution), $n=-1$ (de Sitter solution), $n=-3$ (Einstein
static universe) and $x=const$, $n<-5.83$ or $n>-0.17$ (Klein-Tolman
solution). They appear either as exceptional cases, when equations simpler
than Eq. (2) are to be solved, or as Eq. (2) with $f\left( z\right) =\alpha
z+\beta $, the simplest integrable case. Some candidate $n$ lead to Eq.
(63), but at least one of the constants there has the wrong value.

It has been shown that Eq. (2) stands in the centre of the problem,
independent from the approach or the coordinate system and the integrable
cases may be derived in a unified manner. The problem is a very strange
mixture of simple integrable cases and extremely difficult non-integrable
ones. Perhaps this gives some explanation why the de Sitter and the Einstein
solutions were found in 1917, two years after the appearance of general
relativity, and why the next years have brought only a five-fold discovery
of the KT solution.

An important point is that the problem has been pushed to the mathematical
realm of Abel differential equations and further progress depends on
developments in this field. Using the Buchdahl transformation we have shown
that the case $n=-5$ is integrable and have given its metric and fluid
characteristics. However, its master equation in isotropic coordinates (101)
does not seem to be present in the handbooks. This leaves this paper with an
open end and one may hope that other integrable cases will be found in the
future.

Finally, it is interesting to note that all discrete integrable cases $%
n=-1,-3,-5$ are regular and fall in the interval where the singular KT
solution does not exist. One is tempted to speculate that in this interval
there is a regular one-parameter solution, encompassing the discrete cases.

\end{document}